\documentclass{article}% uncomment this for twocolumn layout and comment line below
\usepackage[utf8]{inputenc} %unicode support
\usepackage{graphicx}
\usepackage{todonotes}
\usepackage{amsmath}
\usepackage{multicol}
\usepackage{comment}
\usepackage{breakcites}
\usepackage{authblk}
\usepackage[symbol]{footmisc}

\title{The Rhythms of the Night: increase in online night activity and emotional resilience during the Spring 2020 Covid-19 lockdown}
\date{} 
\author[1,2]{Maria Castaldo}
\author[2]{Tommaso Venturini}
\author[1]{Paolo Frasca}
\author[3]{Floriana Gargiulo\thanks{Correspondence: floriana.gargiulo@cnrs.fr}}

\affil[1]{Univ.\ Grenoble Alpes, CNRS, Inria, Grenoble INP, GIPSA-lab,
  11 rue des Mathématiques,
  F-38000,
  Grenoble,
  France}
 
\affil[2]{CNRS, CIS-lab,
  59 rue Pouchet,
  F-75017,
  Paris,
  France}
  
\affil[3]{CNRS, Université Paris-Sorbonne - Paris IV, GEMASS,
  59 rue Pouchet,
  F-75017,
  Paris,
  France}

\begin{document}

\maketitle
%\footnote{Correspondence: floriana.gargiulo@cnrs.fr}

%%%%%%%%%%%%%%%%%%%%%%%%%%%%%%%%%%%%%%%%%%%%%%
%%                                          %%
%% Enter the authors here                   %%
%%                                          %%
%% Specify information, if available,       %%
%% in the form:                             %%
%%   <key>={<id1>,<id2>}                    %%
%%   <key>=                                 %%
%% Comment or delete the keys which are     %%
%% not used. Repeat \author command as much %%
%% as required.                             %%
%%                                          %%
%%%%%%%%%%%%%%%%%%%%%%%%%%%%%%%%%%%%%%%%%%%%%%

%%%%%%%%%%%%%%%%%%%%%%%%%%%%%%%%%%%%%%%%%%%%%%
%%                                          %%
%% Enter the authors' addresses here        %%
%%                                          %%
%% Repeat \address commands as much as      %%
%% required.                                %%
%%                                          %%
%%%%%%%%%%%%%%%%%%%%%%%%%%%%%%%%%%%%%%%%%%%%%%

%%%%%%%%%%%%%%%%%%%%%%%%%%%%%%%%%%%%%%%%%%%%%%
%%                                          %%
%% Enter short notes here                   %%
%%                                          %%
%% Short notes will be after addresses      %%
%% on first page.                           %%
%%                                          %%
%%%%%%%%%%%%%%%%%%%%%%%%%%%%%%%%%%%%%%%%%%%%%%

{ \begin{abstract}
\noindent\textbf{Context}. The lockdown orders established in multiple countries in response to the Covid-19 pandemics are arguably one of the most widespread and deepest shock experienced by societies in recent years. %human behaviors in recent years. 
Studying their impact trough the lens of social media offers an unprecedented opportunity {to understand the susceptibility and the resilience of human activity patterns to large-scale exogenous shocks. %In this context, we address two interconnected research questions. 
% On the one hand, 
Firstly, we investigate the changes that this upheaval has caused in online activity in terms of time spent online, themes and emotion shared on the platforms, and rhythms of content consumption. %On the other hand, 
Secondly, %on the grounds that Covid-19 represented a unique and abrupt change on human routines, 
we examine the resilience of certain platform characteristics, such as the daily rhythms of emotion expression.}
\newline
\textbf{Data}. %We base our research on the analysis and comparison of 
{Two independent datasets about the French cyberspace: a fine-grained temporal record of almost 100 thousand YouTube videos and a collection of 8 million Tweets between February 17 and April 14, 2020}. 
\newline
\textbf{Findings}. In both datasets we observe a reshaping of the circadian rhythms with an increase of night activity during the lockdown. The analysis of the videos and tweets published during lockdown shows a general decrease in emotional contents and a shift from themes like work and money to themes like death and safety. However, the daily patterns of emotions remain mostly unchanged, thereby suggesting that emotional cycles are resilient to exogenous shocks. 
\end{abstract}}

\section*{Introduction}
 %\cite{koon,oreg,khar,zvai,xjon,schn,pond,smith,marg,hunn,advi,koha,mouse}
The lockdown established in France from March 17th to May 11th as a response to the Covid-19 pandemic created a sudden and severe transformation of daily routines. This disruption represents a textbook case of exogenous shock on human behaviors, which can, \emph{by comparison}, reveal features of normal social life. In this paper we carry out this comparative analysis, focusing on online behaviors and leveraging a unique YouTube dataset. A few weeks before the lockdown, we had started following a corpus of more than one thousand French political YouTube channels with an exceptional temporal granularity – recording \emph{hour by hour} the number of views of all of their videos. This collection provided us with a unique dataset to study how the lockdown transformed the circadian rhythms of online activities. To make sure our findings are not platform specific, we compare the results obtained on YouTube with a Twitter dataset of 8 million tweets in French.

In this paper, we  investigate changes in the daily rhythm of online activities and address two related research questions: what are the changes produced by the lockdown and how resilient are online circadian rhythms. The analysis of such a exceptional period allows us to distinguish the dynamics due to the Covid-19 crisis from stable features of the social media we investigate.
\begin{comment}
In particular we analyze on a large scale if the lockdown has generated a significant change in the sleep habits in the population and if such change caused any variation of online emotional cycles.

Since circadian online patterns are strongly dependent on users' demography and on platforms' scope, we used two different platforms to conduct our study: YouTube and Twitter. The first one gives us a vision from the standpoint of online objects consumption (video views) the second one from the standpoint of users' production (tweeting and re-tweeting activity). This comparative framework will be central to emphasize which findings are  not platform specific and which ones , on the other hand, are characteristic of a certain platform. 
\end{comment}
%\newline
We focus our research on a single country, France, in order to identify precisely the start and end dates of the lockdown and to work on a (relatively) uniform population sample that excludes national differences.

%\newline
Our analysis shows, for both platforms, an increase of online activities, likely a consequence of the decrease of real-life interactions. The growth in online participation and content consumption is not uniform across the 24 hours, but it is more salient during the night. Besides this variation in volume, we register changes in the kind of content shared from an emotional and thematic point of view. In both platforms, the lockdown is marked by an {unexpected decrease in emotional contents} 
%which may stuns the reader, given the sensitivity of the Covid-19 affected aspects of our life. On a more predictable side, we observed 
{ and by a more predictable thematic shift from topics like "social life" and "leisure", before the lockdown, to matters related to "house" or "death" during the lockdown.}
Against these lockdown-induced changes, some constants of YouTube and Twitter stand out. Despite its quantitative change, the shape of the daily cycle of different emotions (i.e. their prevalence by hour) is not impacted by the lockdown: this finding confirms the results of previous studies that showed a stability of emotional rhythms across seasons and cultures \cite{golder2011diurnal, dzogang2017circadian, lampos2013analysing}. The resilience of these patterns, despite the disruption of the lockdown, {may be due to} biological origins of the emotional expression, which seems to be more influenced by the biological clock than by exogenous factors.

\section*{Related Work}
The patterns of activity in different social media vary according to the characteristics and the scope of the different platforms and to the composition of its user pool. Several papers in the literature have studied circadian fingerprints in different social media. In YouTube, scholars have investigated the rhythms of content creation (i.e. video posting) \cite{szabo2010predicting} and of content fruition (i.e. videos watching) \cite{gill2007youtube}. Daily and a weekly use of Twitter has been analyzed in various countries \cite{ten2014circadian}. Among other platforms, Foursquare has attracted researchers' attention for its circadian and geographic patterns \cite{noulas2011empirical} and its similarities with Twitter \cite{grinberg2013extracting}, while Wikipedia editing patterns have been studied in \cite{yasseri2012circadian} with an interesting focus on inter-cultural variations. Moreover, slightly deviating from the social media framework, mobile phones and instant messaging activities have been frequently analyzed in \cite{pozdnoukhov2010exploratory, jo2012circadian}. Despite their differences, all platforms unsurprisingly show a substantial decrease of activity during the night. 

The realm of night owls is characterized not only by a general quiet but also by typical emotional markers. Emotional content of Twitter activity has been studied in several papers with different approaches~\cite{golder2011diurnal, dzogang2017circadian, lampos2013analysing}. Despite the variety of data considered, all these papers share the same finding: digital nights are consistently characterized by a low level of shared emotions (both positive and negative). 
 
Covid-19 lockdown partially disrupted these rhythms, by drastically changing people's habits and daily routines: from physical commuting to telecommuting, from school classes to distance learning, from in-person meetings to video calls. Within few days whole populations had to re-adapt their behaviors into a new life scenario characterized by deep health and professional concerns. Such unpredictable situation strongly impacted peoples' sleep-wake cycles, as reported in several studies based on surveys like \cite{doi:10.1111/jsr.13119}. 
With our study, therefore, we create a link between the literature on circadian online rhythms and the one on the stress experienced by many societies in recent months. 

\section{Data and Methods}
Since circadian patterns of online activity are strongly dependent on users' demography and on platforms' scope, as mentioned in the introduction, we  conducted our study across two different platforms: YouTube and Twitter. 
This comparative framework will be central to distinguish platform-specific findings from more general trends. 

\subsection{The YouTube dataset}
As mentioned in the introduction, the YouTube dataset is particularly interesting {  
because of its temporal granularity. Indeed, historical data about the number of views can no longer be retrieve from the YouTube application programming interface (API), which only returns the views at the moment of querying and not their temporal evolution. Therefore, studies like the ones by \cite{pinto_using_2013} and \cite{nguyen_analyzing_2019} involving time series of engagement metrics are not longer possible, unless constantly querying the API all along the period of interest, as we did.
} 

{
The dataset covers 1031 popular French channels that (according to the experts we consulted) are particularly influential in the French public debate. These channels, with their description, are listed in the Supplementary Material. The channels have been selected through a qualitative analysis of the French YouTube, aiming to identify a relevant actors that diffuse political opinions through the platform. The selected channels belong to the following categories: local and national media; influential Youtubers discussing political topics; militant associations; politicians; candidates for the European elections; political party; Yellow Vests groups; associations devoted to public causes; large public or private institutions. YouTube provides no information about the location from which videos are viewed, but since the channels of our corpus focus on French public debate, we can assume most of their viewers to be based in France. In collaboration with the Qatar Computing Research Institute (QCRI), we recorded \textit{hour by hour} the evolution of views of each videos published after February 17, for an entire week after the publication. Between February 17 and April 14, we collected the views time series for 99.992 videos. For every video, we also collected the title, the description and other metadata available through the official YouTube API.}

{ Before being able to perform our analysis, some preprocessing turned out to be necessary. Contrarily}
to what one might expect, the total number of views of a given video are not always increasing. Occasionally, YouTube removes from its counting earlier views that it deems to be \textit{fake} because likely produced by bots, click-farming or through other illegitimate tricks, { since YouTube wants "to make sure that videos are viewed by actual humans and not computer programs" \cite{google}, as per the official support web-page}. Since the corrections of these views are made after their recording, our dataset contain hours with a negative number of views. To correct such negative views, we have preprocessed our data in order to uniformly redistribute the corrections made at a given time on the previous hours. 
More precisely, if we call $v_h$ the views collected by a generic video at hour $h$ after publication and $T_h$ the total number of views at hour $h$, if $T_{h+1} < T_{h}$, we correct the time series as follows:
\[
\hat T_{j} = \left(1 - p\right) T_{j}  \;\;\;\;\;\; j  = 1, \ldots, h
\]
where $\hat T_{j}$ is the corrected time series until hour $h$
and $p = \frac{T_{h}-T_{h+1}}{T_{h}}$ is a percentage of correction.
\newline

\subsection{The Twitter dataset}
The Twitter dataset comprises about 8 millions tweets, { retrieved through the official Twitter API,} posted by 5161 {active but} non-professional users from February 17 to April 14. 
{ These non-professional users have been identified within a wider dataset of about 33 millions tweets containing Covid-19-related content.} 
This corpus of tweets was collected by Science–Po MediaLab in Paris, by using the python based scraper Gazouilloire \cite{mediaLab}, a tool developed by Dime Web for systematic and configurable Twitter data collection through Twitter’s official API. The data were collected based on a query of Covid-19-related words in French. All twitting and re-twitting times were collected in European Central Timezone (UTC +1).
 { Being Covid-19 the most trending communication topic in France (as we can observe for example from Google Trends), since its first diffusion in Europe and even more around the lockdown decision, we can assume that a relevant part of the French Twittosphere is potentially present in this database. }

{ Differently from the YouTube database, which does not allow extracting information on the users, for the Twitter database we can define precise profiles of the users we are interested to study, in order to have a more homogeneous, even if reduced, population.}
Since we are interested in the activity of common Twitter users, we decided to exclude newspapers, bloggers, radios, associations, etc. and only consider non-professional users. { Moreover, recalling that the aim of this work is to provide robust cross-platform results, we needed the Twitter dataset to be comparable to the YouTube one: we hence had to restrict the content production on Twitter to the sole production taking place in France. The requirements listed above, together with the filtering they set off, can be clarified as follows:} 
\begin{itemize}
\item facing the need to exclude professional users, we only considered those (1) whose profile did not contain keywords associated to professional use of the platform (e.g. "media", "blog", "official", etc.); (2) with a number of followers lower than the median of the whole dataset; (3) with an activity lower than the median activity ($\sim$400 tweets by week);
\item targeting a significant statistical analysis, we discarded users who published less than 100 tweets in the whole period;
%{\color{blue} some comments on why?}
\item focusing on France, we filtered the sole users who explicitly declared their location to be in France, searching in the location descriptions multiple translations of "France", the name of all the cities in France with a population larger than 10000 inhabitants (with eventual translations), the French regions and departments.
\end{itemize}
{ This filtering left us with 5161 users: their Twitter profiles' descriptions have been manually checked to be sure that the non-professional selection mechanism was effective. Finally, in order to construct the dataset of tweets, the entire timelines of the selected users have been collected through the Twitter API, including tweets not related to Covid-19. Even though this sample is not meant to be representative of the full French Twittosphere, it provides a complete perspective of a sizable and homogeneous set of active non-professional users of the platform over the relevant time frame.}

\subsection{The tool for emotion and topic mining}
To analyze the emotional and thematic contents of tweets and YouTube videos, we used a very well-known and tested tool: the LIWC dictionary \cite{tausczik2010psychological}. The LIWC dictionary classifies words on more than 70 emotional, stylistic and thematic dimensions and has been used in several analogous studies such as \cite{golder2011diurnal, dzogang2017circadian}. { More precisely, the LIWC dictionary provides a list of words associated to each of the dimensions (for example, the category "Positive emotions" contains words like: happy, pretty, and good.} Since the texts we analyze are written in French, we used the French version of the LIWC dictionary \cite{piolat2011version}. 

\bigskip
\section{Results}
\begin{figure*}[h!]
\includegraphics[width=0.98\textwidth]{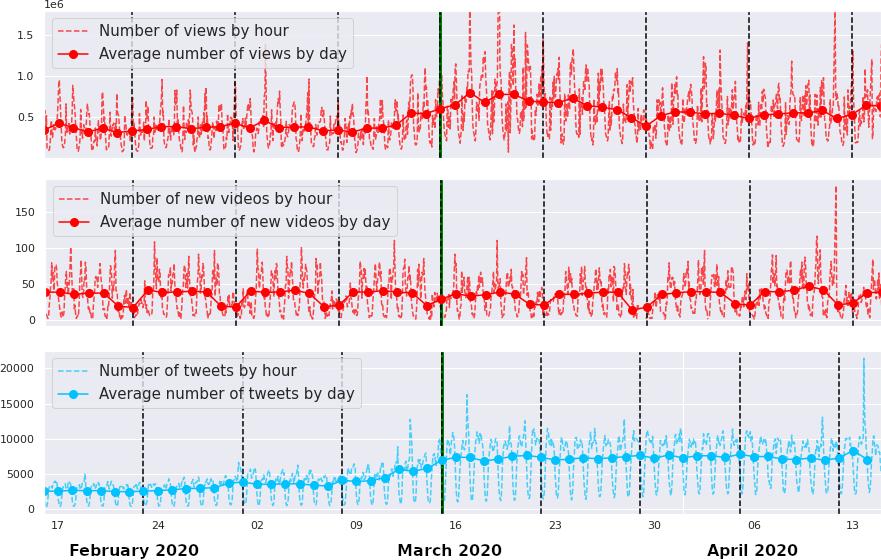}
\caption{%\csentence
{Increase of platforms activity after the lockdown.} 
(top)
Evolution of number of visualizations by day on YouTube. (middle) Evolution of number of new published videos on YouTube. (bottom) Evolution of number of Covid-19 related tweets or re-tweets}
\label{fig:totIncrease}
\end{figure*}

{ The first major habit change generated by the 2020 spring lockdown in France is a considerable increase in the online activity. Recalling that the French first lockdown was announced on March 15 and enforced on the 17th, we can better understand the time series for Twitter posting, YouTube posting and YouTube views displayed in Fig.~\ref{fig:totIncrease}}. All three time series show weekly and daily fluctuations. The average daily signals, smoothed by a moving average over a 7 day rolling window, reveal an increase of activity for  Twitter posting and YouTube watching around the beginning of the lockdown. { Notice that on both platforms the increase of activity started from the very moment the lockdown was announced (which itself sparked much debate)}.  

As for the  posting of videos on YouTube, such an activity is less casual and more stable than tweeting (particularly for the high-visibility channels that we monitored) and therefore conserved  the same weekly and daily rhythms during the lockdown. 
Since the video production is hardly affected by the lockdown, we will not consider this dimension in the rest of the paper.  

\subsection{The rhythm of the night}
\begin{figure*}[h!]
\includegraphics[width=0.98\textwidth]{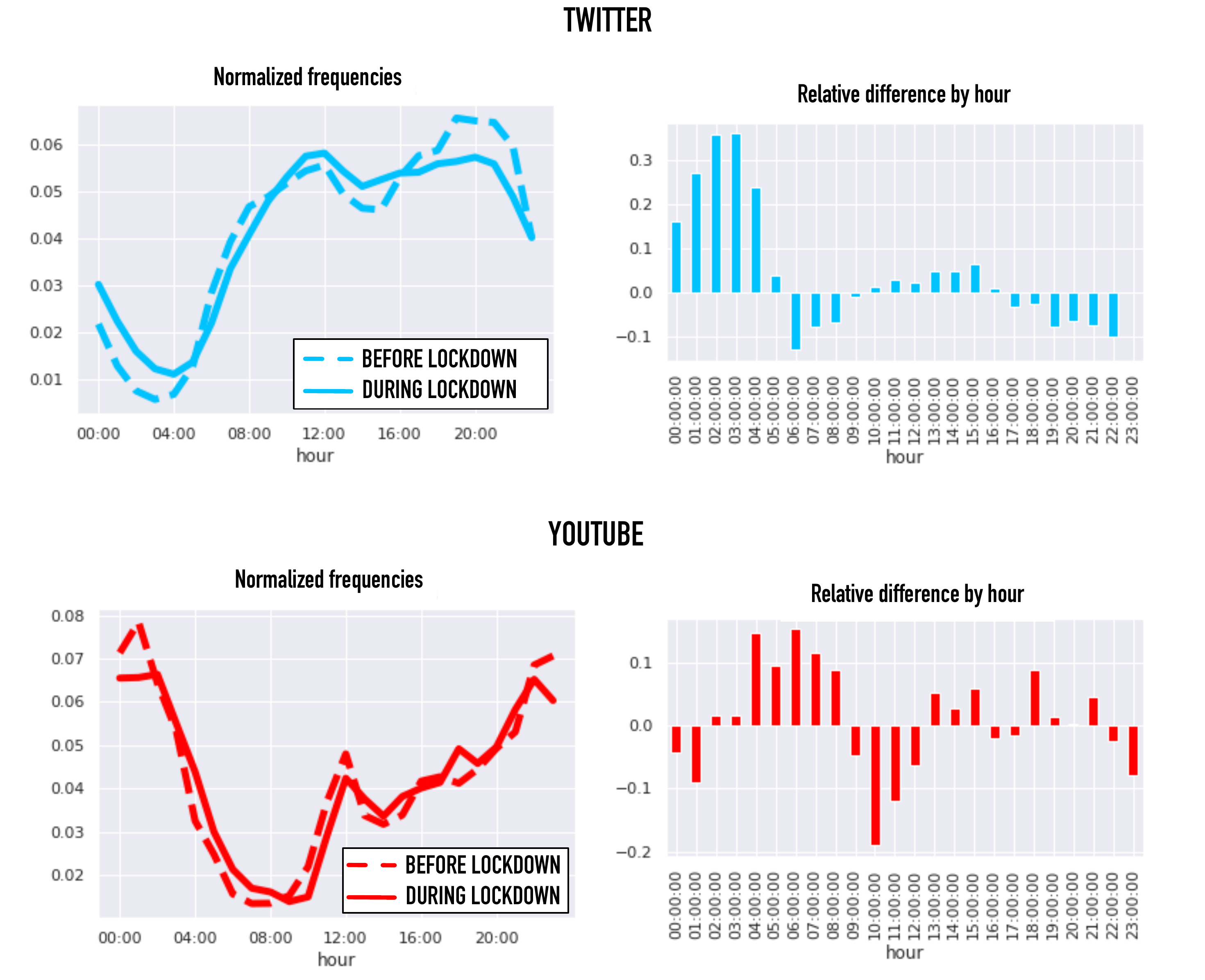}
\caption{%\csentence
{Circadian Rhythm Changes.}
On the left, Twitter and YouTube circadian rhythms before and during the lockdown. On the right, we explicitly evaluate the relative differences between rhythms before and during the lockdown}
\label{fig:RelativeIncrease}
\end{figure*}

To highlight the effects of the lockdown, in the following we analyze thematic and emotional changes before and after the start of the lockdown (March 17). We refer to the period \textit{before} the lockdown as to the three weeks from February 17 to March 9. At the same time we will refer to the period \textit{after} the lockdown enforcement as to the three weeks from March 23 to April 14. To exclude the transient effects of the transition phase, we discard the data about the two weeks around the lockdown onset (from the 9th to the 22nd of March). Preliminarly, we calculated the normalized daily activity profiles before and during lockdown for each hour of the day by 
\begin{equation}
    f^{\text{Twitter}}(h)=\frac{\sum\limits_{d \in \text{days}} N_{\text{tweets}}(d,h)  } {\sum\limits_{d \in \text{days}}\sum\limits_{h \in \{0,\ldots,23\}} N_{\text{tweets}}(d,h)  }
\end{equation} and \begin{equation}
    f^{\text{YouTube}}(h)=\frac{\sum\limits_{d\in \text{days}}N_{\text{views}}(h)  } {\sum\limits_{d\in \text{days}}\sum\limits_{h \in \{0,\ldots,23\}} N_{\text{views}}(d,h),  }
\end{equation}
where $h \in  \{0,\ldots,23\}$ are the hours of day, $d$ are the days considered, and $N_{\text{tweets}}(d,h)$ and $N_{\text{views}}(d,t)$ are respectively the number of tweets and of YouTube views at hour $h$ of day $d$.
The results are reported in the left plots of Fig.~\ref{fig:RelativeIncrease}. We first observe that the profiles for Twitter and YouTube are quite different: while Twitter is mostly used during the day, with a strong activity decrease after midnight, YouTube is characterized by a higher night activity.  While Twitter is an active media, characterized by a debating and prosuming culture \cite{Ritzler} that encourages participation at the time of the day when engagement is maximum, videos watching on YouTube is, for many  users, a more passive activity \cite{Khan} which can easily fit the more relaxed late hours.

To quantify the differences between profiles before and during the lockdown, we calculated the relative differences of the normalized profiles:
\begin{equation}
    \delta (h)=\frac{f^{\text{after}}(h)-f^{\text{before}}(h)}{f^{\text{after}}(h)+f^{\text{before}}(h)}
\end{equation}
This quantity is reported in the right plots of Fig.~\ref{fig:RelativeIncrease}. Both  YouTube and Twitter experienced an  activity increase during the night and a smaller decrease of the activity in the early morning (6am-9am for Twitter and 9am-12am for YouTube). We observe that the morning decrease in Twitter is much smaller than the night increase. This suggests that, with the lockdown, people stayed longer awake during the night but without oversleeping in the morning. 

\begin{figure*}[h!]
\includegraphics[width=0.98\textwidth]{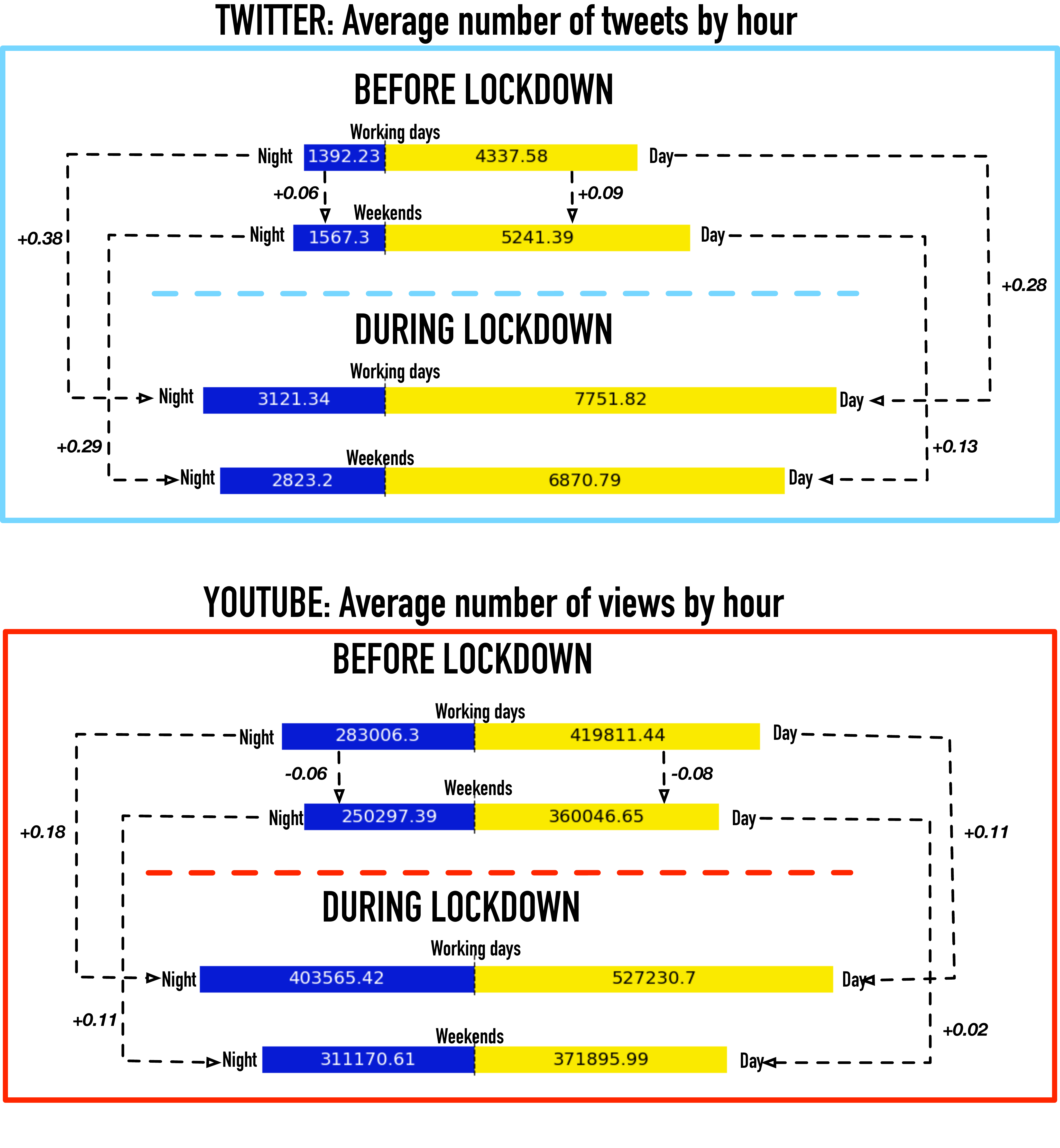}
\caption{%\csentence
{Night-vs-day and working day-vs-weekend patterns.}
Average number of Tweets (upper plot) or YouTube views (lower plot) by hour, during day and night, working days and weekends. The numbers next to the dotted lines represent the relative increment between the related quantities. }
\label{fig:weekend}
\end{figure*}

{ To appraise the variations in the activity during lockdown nights, we compare this variation with another factor known to impact online circadian patterns: the weekly cycle of weekends and working days. We aggregate the data at the level of day and night: based on the shapes of the curves in Fig.~\ref{fig:RelativeIncrease}, we consider night on Twitter the hours between 11PM and 6AM of the following day and on YouTube the hours between 1AM and 8AM. We decompose the aggregated activity into weekends and working days. To obtain comparable measures we divide the activity counts by the number of hours of the corresponding time period (night=7, day=17) and by the number of days of week parts (weekend=2, working days=5). The average number of tweets/views in the different categories are represented in Fig.~\ref{fig:weekend}, together with the relative changes among the classes. As we already observed, the night variations are the largest changes, both in weekends and working days. The activity increments due to lockdown are significantly larger than the variations associated with the normal the week cycle. In YouTube the variations also have opposite signs: while the activity normally decreases during weekend, it increases for lockdown. For both the platforms, and both for day and night, the most significant difference concerns the working days:  the augmented social media usage seems to be replacing the time previously dedicated to daily routines (like commuting to work, going to sleep early, etc.) more than the time of recreational activities.}

\begin{figure*}[h!]
\includegraphics[width=0.98\textwidth]{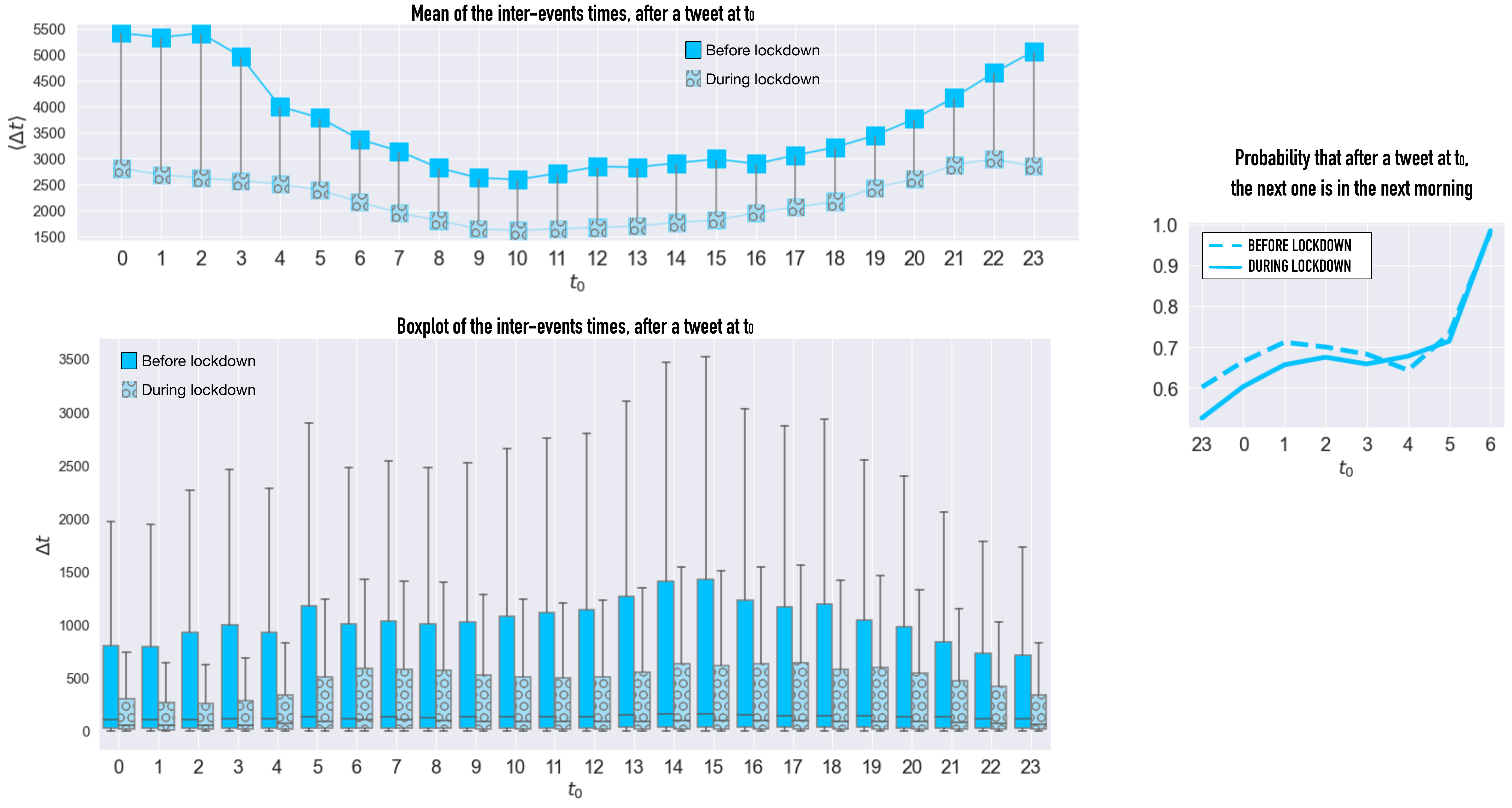}
\caption{%\csentence
{Interevents.}
Left plots: Average inter-event distance for an event starting at time $t_0$ and boxplot of the interevents starting at $t_0$. Right plot: Probability that a tweet following a tweet at $t_0$ is posted on the following morning (dashed=before lockdown, solid=during lockdown).}
\label{fig:interevents}
\end{figure*}

To confirm the hypothesis of reduced sleep during the lockdown, we analyze the situation at the individual level. For each Twitter user, we calculate the average time lag between two consecutive Tweets. Fig.~\ref{fig:interevents} shows the hourly average of this measure {color{red} (and the boxplots to appreciate its variability) before and during the lockdown}. While, in normal times, the average inter-event times are much higher during the night (because of sleeping breaks), the  lockdown flattened the curve, thereby suggesting a shortening of sleep intervals \cite{doi:10.1111/jsr.13119}. { In the right plot of Fig.~\ref{fig:interevents} we display, for each hour of the night, the probability that the subsequent tweet is posted the following morning (between 7 AM and 12AM) rather than during the same night. This quantity represents the probability to "go to sleep" after an event at $t_0$. For this analysis we only considered the inter-events that are longer than 1 hour, in order to get the last action only in a potential activity burst and reduce the noise. We also excluded all the inter-events finishing after the next morning. We see that, until 4AM, the probability to get asleep is lower during lockdown.
%\footnote{
{ These findings are based on an aggregated statistical study of the inter-events. Unfortunately, the short time span of our data collection (2 months) does not allow a more sophisticated analysis, for example, of the users chronotypes, which, as shown by \cite{aledavood2018social, aledavood2015daily} can be extremely important for understanding the individual sleep behavior.}
%}%end footnote
} % end of red

\subsection{What is night?}
\begin{figure*}[h!]
  \includegraphics[width=1.08\textwidth]{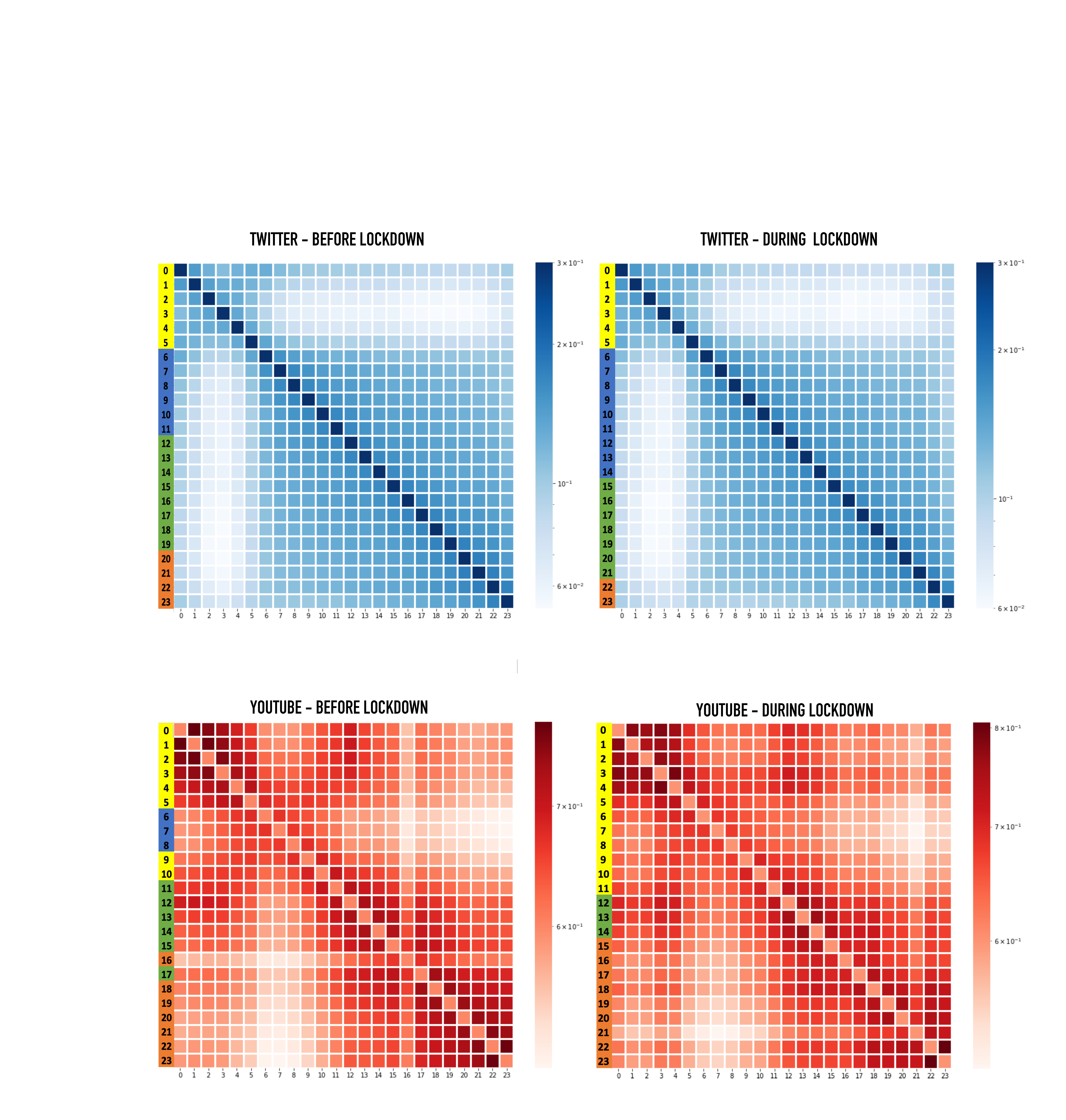}
  \caption{%\csentence
  {Hours Correlation.}
Heat-maps of the content hour similarity before and during the lockdown for Twitter and YouTube. The colors of the hours on the left of the graph represent the partition of the k-means clustering.}
 \label{fig:correlationMap}
\end{figure*}

In this paragraph we analyze whether the quantitative changes observed in the previous paragraph correspond to differences in terms of contents. For Twitter, we build the sets $K(d,h)$ containing all the hashtags posted in day $d$ at hour $h$. For YouTube, we build the sets $K(d,h)$ containing all the videos visualized in day $d$ at hour $h$. We define the time similarity matrix, $\Theta$, between two day's hours $h_1,h_2$, based on the Jaccard similarity between the sets $K(d,h_1)$ and $K(d,h_2)$ as:
\begin{equation}
    \Theta(h_1,h_2)=\frac{1}{N_{\text{days}}}\sum_{d\in \text{days}}J(K(d,h_1),K(d,h_2))
\end{equation}
where $J(K(d,h_1),K(d,h_2)) = \frac{|K(d,h_1)\cap K(d,h_2)|}{|K(d,h_1)\cup K(d,h_2)|}$ is the Jaccard similarity.
Matrix $\Theta$, represented in Fig.~\ref{fig:correlationMap}, indicates how the content shared or viewed at a certain hour is similar to the content in all the other hours. 
We perform a $k$-mean clustering procedure on this matrix to better identify the relationships between the hours of the day. { The results of the clustering are represented by the colors of the hours on the left of the plots of Fig.~\ref{fig:correlationMap}. Remarkably, these clusters identify different periods of the day, thereby showing that contents evolve along the day and different periods of the day are characterized by specific contents.
Both for Twitter and YouTube, and both before and during lockdown, night hours (0am-5am) are characterized by contents distinctively different from the rest of the day. Before the lockdown, morning hours (6am-10am for Twitter and 6am-8am for YouTube) were the most different from night-time and constitute a well definite cluster. For Youtube before the lockdown, the late morning hours 9am-10am showed a return of the night contents. For both platforms before the lockdown, we observe a lunch-afternoon cluster and an evening cluster. Lockdown affected the morning cluster, though in opposite ways for the two platforms. In Twitter, the morning cluster extended to first hours of the afternoon cluster (lunch hours), while in YouTube the night vibe extended into the morning (until 11am).
In Twitter, the lockdown shifted the afternoon cluster after lunch time (3pm-9pm) and consequently reduced the evening hours (10pm-11pm).  In YouTube, we observe the emergence of contents dedicated to the lunch hours (12am-2pm) and a second uniform block until covering the afternoon and the evening (3pm-11pm).}

\subsection{A significant change of content}
\begin{figure*}[h!]
  \includegraphics[width=0.98\textwidth]{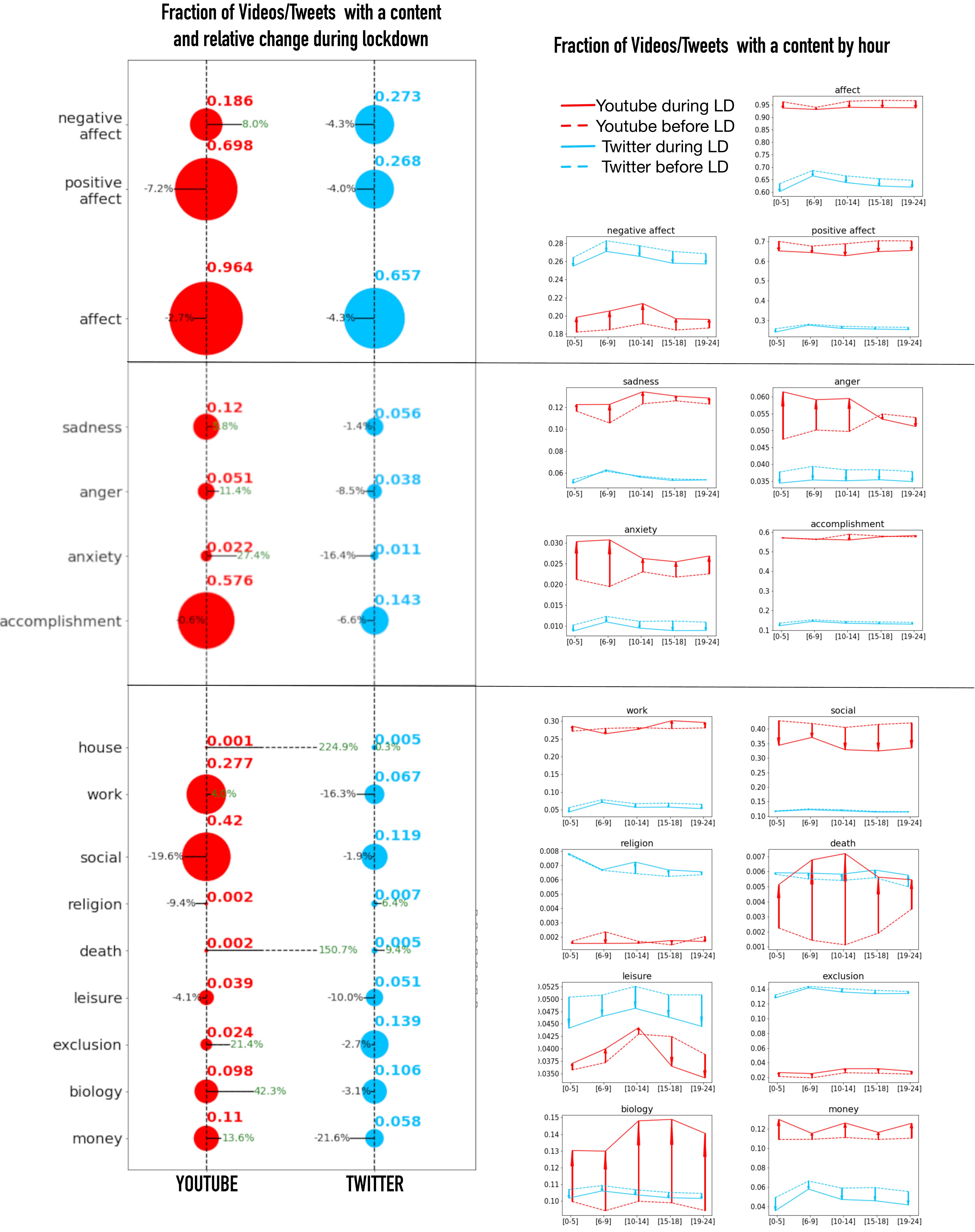}
  \caption{%\csentence
  {Themes and Emotions before and during lock-down.}
 Left plot: Fraction of videos/Tweets with a content and relative change with lockdown. The size of the points is proportional to the fractions (quantified by the upper numbers). The orientation of the line indicates if there was an increase (orientation toward right) or decrease (toward left) with the lockdown. The length of the line is proportional to the percentage increase/decrease with the lockdown. 
Right plot: Fraction of videos/Tweets with a content by hour. The continuous line indicates the fractions after the lockdown, the dotted lines before. An arrow starts from the before to the after line for each hour period: if the arrow is oriented towards the top, the lockdown increased the content fraction in the selected hours and viceversa. 
In both plots YouTube is in red, Twitter is in blue. }
\label{fig:emotions}
\end{figure*}

We proceed with the analyses of the emotional and thematic content of Twitter and YouTube activities, before and during the lockdown. 
Using the categories of the LIWC dictionary we will consider three separated analytic dimensions: 
\begin{itemize}
    \item General Affects: Positive Affect, Negative Affect;
    \item Specific Emotions: Sadness, Anger, Anxiety and Accomplishment:
    \item Thematic contents: Work, Social life, Religion, Death, Fun, Exclusion, Biology, Money.
\end{itemize}
We first assign the tweets and YouTube videos (based on the words contained in their titles and descriptions) to one category for each dimension. To do so, we count how many terms from each category are contained in each tweet/video, and we assign the content to the prevalent category. For example, for the dimension "General Affect" each content is categorized as either Positive or Negative Affect or not classified if the items contains no categorical words or similar proportions of positive and negative terms. We also consider the global emotional level ("Affect") of the items, by counting together the positive and negative words. 
For each dimension, we compute the fraction of tweets and retweets in each category during the lockdown and the difference compared to the previous period. In the same way, for YouTube, we evaluate the fraction of visualization of videos in each category over the total number of views. The results are reported in the left plot of Fig.~\ref{fig:emotions}. 

Comparing the two platforms, we first observe that YouTube is more "emotional" than Twitter and is generally populated by more positive content. Both platforms experience a decrease of the emotional sphere during the lockdown, { but on YouTube in particular we observe an increase of emotionally negative contents.}
Regarding specific emotions, we notice that expressions of accomplishment decline in both platforms. Instead, while Twitter experienced a decrease of all specific emotions, YouTube, which was already characterized by a higher level of anger, sadness and anxiety before the lockdown, goes through an important increase of these sentiments. 
From a thematic point of view we observe, unsurprisingly, a decrease of the contents related to {social life and leisure} and an increase of contents related to death and house. On Twitter we also have a significant increase of religion-related contents. 
All differences in distribution of contents before and during the lockdown have been tested statistically with Kolmogorov-Smirnov (KS) tests. For both platforms before and during the lockdown an hourly aggregation has been performed in order to get two different samples of emotion distribution. The results of the KS test over those distributions are reported in Table 1 and Table 2.
and most of the times support the statistical relevance of differences in content distribution before and during the lockdown. 

\subsection{Because the night belongs to...}
Drawing on our previous hour clustering, we divide the day according to five time periods: [0am-5am],[6am-9am],[10am-2pm],[3pm-6pm],[7pm-12pm] to identify a circadian profile for each of our categories (right plot of Fig.~\ref{fig:emotions}). For several categories related to emotions, we can first notice an interesting Twitter/YouTube difference: what peaks in the early morning on Twitter [6am-9am] tends to peak in the following interval in YouTube [10am-2pm], thereby suggesting that YouTube content is consumed later in the day. Going into more detail, in agreement with the findings of \cite{golder2011diurnal, dzogang2017circadian, lampos2013analysing} we observe that nights are characterized by low emotional levels, especially positive ones, while both positivity and negativity tend to peak at the moment of the awakening. This pattern is more evident on Twitter also at the level of specific emotions, while on YouTube a significant portion of negative contents is consumed during the night.

An interesting observation, again in phase with the precedent findings of \cite{golder2011diurnal, dzogang2017circadian, lampos2013analysing}, is that the daily emotional patterns seem to be resilient to the Covid-19 disruption: even if the volumes of some emotions changed during the lockdown, their daily distribution generally maintained the same shape, as demonstrated by a rough parallelism of the lines before and during the lockdown (with some exceptions regarding anger and anxiety on YouTube). This fact confirms the observation made in \cite{golder2011diurnal} that external factors, even as important as the Covid-19 lockdown, influence the emotional patterns less than the sleep-wake cycles. Interestingly, this is not the case for the distribution of topics which has been more significantly influenced by the lockdown. 
\\

\section{Discussion}
The Covid-19 pandemic and the ensuing lockdown have deeply and widely disrupted people's daily routines. Our research exposed some of these changes through the lens of social media. By highlighting what has changed during the lockdown and what has resisted the Covid-19 shock, our research proves that certain online habits are more resilient to external disruption than others, thereby suggesting which human behaviors are more influenced by exogenous factors and which are, on the contrary, constant even in exceptional situations.

Circadian rhythms of activity proved to be strongly related to lifestyle and working hours: the change observed in the lockdown rhythms suggests in particular that in the absence of external constraints such as school and office hours, the boundaries between night and day become more flexible. 

Perhaps surprisingly, we found out that the emotional charge of tweets and YouTube videos decreased with the lockdown, arguably leaving space to less emotional contents. However, this general finding is tempered by the fact that on YouTube (which is inherently a more emotional medium than Twitter) negative sentiments like anger and anxiety did increase, thus revealing the stressful situation for the population. As topics are concerned, online discussions proved to follow real world events and, unsurprisingly in a global epidemic that forced people at home, shifted towards questions connected to biology, house, and death.

As an even more interesting result, we pointed out the resilience of emotional patterns in online activities. While the general emotional charge of online contents decreased during the lockdown, it maintained its normal daily distribution. After March 17th, nightlife continues to be characterized by less emotional content despite the stress caused by the Covid-19 crisis. Even if the circadian rhythms change and people stay awake longer, at night, they seem to continue to share and consume the same type of contents. 

In future research, we would like to investigate more in depth the nature of this low-emotional nighttime space, in order to reveal weather it consists of more informative and impartial contents and it gives rise to positive and constructive forms of debate or, on the contrary, whether it is more markedly affected by fake news and other types of misinformation. The LIWC dictionary used so far allowed us to distinguish emotional from unemotional contents: more specific dictionaries could lead to deeper insights about nighttime activity and to better understand the relationship between information and emotion in different medias. Broadly speaking, we hope that noticing the resilience of some online patterns during the lockdown might encourage future research in emotional rhythms and their reaction to external shocks.

{ Before concluding, we would like to briefly discuss two limitations of our study. 
In this research, we analyzed two different platforms, therefore providing insights on two different types of public
%: the indistinguishable consumers of YouTube videos and a quite homogeneous set of active Twitter users
. However, we should remind that these two viewpoints are not sufficient to completely represent the French social media and, even less, French society. A comparison with other type of data, not available to our team, like phone call datasets or mobile usage information would be useful to confirm and refine our findings.}\\ 
{
As a second limitation, our analysis is limited to a single country and on the period of the first lockdown. The continued and worldwide impact of the Covid-19 crisis call for a comparison with similar analyses in other countries and in other phases of the pandemics. % would be a necessary further step.
}

%%%%%%%%%%%%%%%%%%%%%%%%%%%%%%%%%%%%%%%%%%%%%%
%%                                          %%
%% Backmatter begins here                   %%
%%                                          %%
%%%%%%%%%%%%%%%%%%%%%%%%%%%%%%%%%%%%%%%%%%%%%%

\section*{Availability of data and material}
Authors agree to make their data available upon request. The list of the YouTube channels is reported in the Supplementary Material. 

\section*{Competing interests}
 The authors declare that they have no competing interests.

\section*{Author's contributions}
%All the authors conceived the idea. M.C. and F.G. analyzed the data and collected the missing data. All authors wrote the manuscript.
M.C. and F.G. collected the missing data and analyzed the data. All authors conceived the research, discussed the results and wrote the manuscript.

\section*{Funding}
This research has been supported by the French national science foundation (Agence Nationale de la Recherche) through  project TRACTRUST and by CNRS through the 80 PRIME MITI project ``Disorders of Online Media'' (DOOM).

\section*{Acknowledgments}
F.G. thanks Jeremy Ward for the useful discussions.  A note of thanks and major acknowledgment goes to Yoan Dinkov and Preslav Nakov of the Qatar Computing Research Institute (QCRI) for their help in collecting YouTube data. A special thanks goes to Bilel Benbouzid for the precious discussions on the French YouTube landscape.  
%%  After submission of the .TEX file,                     %%
%%  you will be prompted to submit your .BBL file.         %%

% if your bibliography is in bibtex format, use those commands:

%\nocite{*}
\small
%\bibliographystyle{bmc-mathphys} % Style BST file (bmc-mathphys, vancouver, spbasic).
%\bibliography{bmc_article}      % Bibliography file (usually '*.bib' )


% for author-year bibliography (bmc-mathphys or spbasic)
% a) write to bib file (bmc-mathphys only)
% @settings{label, options="nameyear"}

% or include bibliography directly:
% \begin{thebibliography}
% \bibitem{b1}
% \end{thebibliography}

%%%%%%%%%%%%%%%%%%%%%%%%%%%%%%%%%%%
%%                               %%
%% Tables                        %%
%%                               %%
%%%%%%%%%%%%%%%%%%%%%%%%%%%%%%%%%%%

%% Use of \listoftables is discouraged.
%%
\clearpage
\section*{Tables}
%\vspace{40pt} %pt?
\begin{table}[h!]
\label{tab:kstest_YT}
\caption{p-values of Kolmogorov-Smirnov tests for YouTube content differences displayed by Fig.~\ref{fig:emotions}, i.e. content differences before and during the lockdown}
\begin{tabular}{l|lll}
\hline
Category                 & \begin{tabular}{@{}c@{}}Avg Before \\ Lockdown\end{tabular} & \begin{tabular}{@{}c@{}}Avg During \\ Lockdown\end{tabular} & KS p-value \\ \hline
Negative Affect & $0.187$                        & $0.202$                       & $2.29\cdot 10^{-9}  $          \\
Positive Affect & $0.693$                        & $0.647$                       &$ \sim 0  $                 \\
Affect          & $0.961$                        & $0.938$                       & $\sim 0$                   \\
Sadness         & $0.118$                        & $0.127$                       & $7.56\cdot 10^{-8}$            \\
Anger           & $5.17\cdot 10^{-2}$                     & $5.78\cdot 10^{-2}$                    & $3.16\cdot 10^{-12}$          \\
Anxiety         & $2.15\cdot 10^{-2}$                     & $2.88\cdot 10^{-2}$                    & $2.31\cdot 10^{-7}$            \\
Accomplishment  & $0.575$                        & $0.571$                       & $0.075$               \\
Work            & $0.278$                        & $0.285$                       & $0.0025$              \\
Social Life     & $0.419 $                       & $0.341$                       & $\sim 0$                   \\
Religion        & $1.91\cdot 10^{-3}$                     & $1.67\cdot 10^{-3}$                    & $1.29\cdot 10^{-4}$            \\
Death           & $1.96\cdot 10^{-3}$                     & $5.64\cdot 10^{-3}$                    & $\sim0$                   \\
Leisure         & $3.95\cdot 10^{-2}$                     & $3.77\cdot 10^{-2}$                    & $1.52\cdot 10^{-10}$            \\
Exclusion       & $2.35\cdot 10^{-2}$                     & $2.85\cdot 10^{-2}$                    & $1.82\cdot 10^{-14}$           \\
Biology         & $9.70\cdot 10^{-2}$                     & $0.14$                        & $\sim 0$                   \\
Money           & $0.109$                        & $0.123$                       & $2.22\cdot 10^{-16}$         \\ \hline  
\end{tabular}
\end{table}

\vspace{10pt}
\begin{table}[h!]
\label{tab:kstest_TW}
\caption{p-values of Kolmogorov-Smirnov tests for Twitter content differences displayed by Fig.~\ref{fig:emotions}, i.e. content differences before and during the lockdown}
\begin{tabular}{l|lll}
\hline
Category                 & \begin{tabular}{@{}c@{}}Avg Before \\ Lockdown\end{tabular} & \begin{tabular}{@{}c@{}}Avg During \\ Lockdown\end{tabular} & KS p-value \\ \hline
Negative Affect & $0.275$                                                                                                                                  & $0.261$                                                                                                                                  & $3.11\cdot 10^{-15}  $ \\
Positive Affect & $0.268$                                                                                                                                  & $0.256$                                                                                                                                  & $ 3.11\cdot 10^{-15} $ \\
Affect          & $0.659$                                                                                                                                  & $0.628$                                                                                                                                  & $3.11\cdot 10^{-15}$   \\
Sadness         & $5.55\cdot 10^{-2}$                                                                                                                      & $5.54\cdot 10^{-2}$                                                                                                                      & $0.984$                \\
Anger           & $3.92\cdot 10^{-2}$                                                                                                                      & $3.50\cdot 10^{-2}$                                                                                                                      & $3.11\cdot 10^{-12}$   \\
Anxiety         & $1.15\cdot 10^{-2}$                                                                                                                      & $9.39\cdot 10^{-3}$                                                                                                                      & $3.11\cdot 10^{-12}$   \\
Accomplishment  & $1.46\cdot 10^{-1}$                                                                                                                      & $1.32\cdot 10^{-1}$                                                                                                                      & $3.11\cdot 10^{-12}$   \\
Work            & $6.96\cdot 10{-2}$                                                                                                                       & $5.58\cdot 10{-2}$                                                                                                                       & $0.0025$               \\
Social Life     & $0.119 $                                                                                                                                 & $0.117$                                                                                                                                  & $1.46 \cdot 10^{-6}$   \\
Religion        & $6.87\cdot 10^{-3}$                                                                                                                      & $7.14\cdot 10^{-3}$                                                                                                                      & $3.85\cdot 10^{-3}$    \\
Death           & $5.51\cdot 10^{-3}$                                                                                                                      & $5.90\cdot 10^{-3}$                                                                                                                      & $1.03 \cdot 10^{-3}$   \\
Leisure         & $5.34\cdot 10^{-2}$                                                                                                                      & $4.57\cdot 10^{-2}$                                                                                                                      & $3.11\cdot 10^{-15}$   \\
Exclusion       & $1.37\cdot 10^{-1}$                                                                                                                      & $1.34\cdot 10^{-1}$                                                                                                                      & $1.73\cdot 10^{-8}$    \\
Biology         & $1.06\cdot 10^{-1}$                                                                                                                      & $1.03 \cdot 10^{-1}$                                                                                                                     & $4.20 \cdot 10^{-8}$   \\
Money           & $5.91 \cdot 10^{-2}$                                                                                                                     & $4.51 \cdot 10^{-2}$                                                                                                                     & $3.11\cdot 10^{-15}$   \\ \hline
\end{tabular}
\end{table}

\newpage
%%%%%%%%%%%%%%%%%%%%%%%%%%%%%%%%%%%
%%                               %%
%% Figures                       %%
%%                               %%
%% NB: this is for captions and  %%
%% Titles. All graphics must be  %%
%% submitted separately and NOT  %%
%% included in the Tex document  %%
%%                               %%
%%%%%%%%%%%%%%%%%%%%%%%%%%%%%%%%%%%

%%
%% Do not use \listoffigures as most will included as separate files

\section*{Figures}

\end{document}